\documentclass[aps,twocolumn,showpacs]{revtex4}
\usepackage{graphicx}
\usepackage{textcomp}

\begin{document}

\title{Acoustic Emission and Shape Memory Effect in the Martensitic Transformation }
\author{S.Sreekala$^{1}$ and G.Ananthakrishna$^{1,2}$}
\affiliation{$^1$ Materials Research Centre, Indian Institute of Science,
Bangalore-560012, India\\
$^{2}$ Centre for Condensed Matter Theory, Indian Institute  of Science, Bangalore-560012, India}

\begin{abstract}
Acoustic emission signals are known to exhibit a high degree of reproducibility
in time and show correlations with the growth and shrinkage of martensite
domains when athermal martensites are subjected to repeated thermal cycling in a restricted
temperature range. We show that a recently introduced two dimensional
model for the martensitic transformation mimics these features.
We also show that these features are related to the shape-memory effect where
near full reversal of morphological features are seen under these thermal
cycling conditions.
\end{abstract}
\maketitle
Acoustic emission(AE) is an important nondestructive technique that is sensitive to the
microstructural changes taking place inside the system.
It is widely employed in the detection of
earthquakes of small magnitudes \cite{Aber}, understanding and mapping
nucleation events of fracture in seismology \cite{Lock}, and precursor effect of
fracture \cite{Garci}. Its nondestructive character finds many applications in
industry as well \cite{Marge}. Acoustic emission is  attributed
 to the sudden release of stored strain energy, although the details of the
 mechanism are generally system specific. For instance, the difference between the nature of
 AE signals  in thermal and athermal transformations has been used to characterize the nature of a
 first order transformation \cite{Riche}.
 In the case of athermal martensites, the kinetics
 of the transformation  is not controlled by thermal fluctuations
 which implies  that the  system is caught in a local minimum. It
  jumps over the barrier only when subjected
  to  changes in  external parameters such as temperature or
stress \cite{Riche,Lovey99,Carr,Picor,Vives1,Vives2}.
 The AE emitted during martensitic
transformation often shows {\it unusual and interesting} statistical properties,
one that has
attracted considerable attention is the power law statistics of the AE signals
\cite{Vives1,Carr,Rajeev}, as it is  reminiscent of self-organized criticality
\cite{Bak}. Much less known, but perhaps equally intriguing
is that AE signals can show a high degree of statistical correlations in time
(reproducibility) and its correlations to the growth and shrinkage of
microplates \cite{Lovey99,Amen,Picor}. These results based on
careful experiments on single crystals of $CuAlZn$ alloys under repeated thermal
cycling in a restricted domain of temperatures \cite{Amen,Picor} are
well summarized in Figs. 10-12 of Ref. \cite{Lovey99}.
{\it To the best of our knowledge,
 these results have not been explained so far}.
Explaining these features is  also crucial to the understanding
of the shape-memory effect of martensites \cite{comment}. Moreover, as far as we
are aware, there are no models of shape-memory effect which demonstrate {\it the
 reversal of morphology of martensite domains under thermal cycling.}
The purpose of this paper is to devise a model which explains these 
correlations and
provide a model for  the shape-memory effect.

Martensitic transformation is considered as an atypical first order
transformation as it exhibits some features of second order transition
such as precursor effect \cite{Kartha} and the scale free power law
distribution of AE signals, again a signature of critical fluctuations.
On cooling from a
higher symmetry parent phase, nucleation of thin plate like product domains with
internal twinned structure is seen. Nucleation  is known to be athermal
usually occurring at defects \cite{Kacha}.
Long range strain fields resulting from the difference in the unit cell structure
between the parent and product phases
tend to block the transformation. The athermal nature of the transformation
also means that the amount of transformed phase is fixed for a fixed quench
and further undercooling is required for an additional growth. For the same reason,
the transformation occurs in a broad range of temperatures and it exhibits
hysteresis under thermal cycling.

The power law statistics of the avalanches has been modeled using random
field Ising models \cite{Sethna} where the extent of 
  quenched disorder plays a crucial role. However,  as the power law is seen
only after repeated thermal cycling, Vives {\it et al}
\cite{Vives1,Vives2} conclude
that the transformation induced disorder, rather than
quenched  disorder ( defects )  is responsible  for these avalanches. Recently,
a model which generates dynamical disorder 
during the transformation has been shown
to reproduce the power law statistics \cite{Rajeev}.
 We  show that this simple 2D
phenomenological model \cite{Rajeev} has
the right ingredients to explain the correlated AE signals and
the shape-memory effect \cite{comment}.

The  model uses the deviatoric strain as the principal order parameter. The
effect of the  bulk and shear strains are included only in a
phenomenologically way through a long range interaction  term (see below).
In terms of the two components of the displacement field, $u_x$ and $u_y$, the
deviatoric strain is given by $\epsilon =(\frac{\partial u_{x}}{\partial x}-
\frac{\partial u_{y}}{\partial y})/\sqrt 2$.

The  scaled free-energy, written entirely in terms of $\epsilon ( \vec r)$,
 is  the sum of local a free energy $F_L$
and an effective  long-range term, $F_{lr}$, that describes transformation induced
strain-strain interaction. $F_L$ is given by
\begin{equation}
F_L  = \int d\vec{r}\bigg[f_l(\epsilon(\vec{r}))+
{\alpha\over{2}}{{(\vec{\nabla}\epsilon(\vec{r}))^{2}}}
-\sigma(\vec{r})\epsilon(\vec{r})\bigg],
\end{equation}
where both $\alpha$ and $\sigma$ are in scaled form.
Here, $f_l(\epsilon(\vec{r}))$ is taken as the usual Landau polynomial
for a first order transition given by
$f_l(\epsilon(\vec{r}))={{\tau}\over{2}}{\epsilon(\vec{r})}^{2}
-{\epsilon(\vec{r})}^{4}+
{1\over2}{\epsilon(\vec{r})}^{6}$, where ${{\tau}=(T-T_c)/(T_0-T_c)}$
is the scaled temperature, $ T_0 $ is the first-order transition
temperature, and $ T_c $ is the temperature below which
there are only two degenerate global minima
$ {\epsilon=  \pm {\epsilon_m} }$. We mimic
the nucleation of the martensite domains occurring at
localized defects sites by an inhomogeneous stress
field, $\sigma(\vec{r})$ \cite{Cao}.
Recent studies have shown
that an effective long-range interaction between the transformed domains
results due to the coupling with the other components of the strain order parameter
\cite{Kartha,Shenoy}. However, as our approach is phenomenological,  we attempt to
guess a simple form that respects the invariance of $f_l(\epsilon)$  under
$\epsilon \rightarrow -\epsilon$. Following Wang and
Kachaturyan \cite{Wang} (see their Eqn (11)), one simple form that satisfies
this requirement when represented in the Fourier space can be written as
\begin{equation}
F_{lr}\{\epsilon\}=\int d\vec{k} B({\vec{k}}/k)
{\{ {\epsilon^{2}}(\vec{r})\}}_{k}
{\{ {\epsilon^{2}}(\vec{r})\}}_{k^{*}},
\end{equation}
where $\{ {\epsilon^{2}}(\vec{r})\}_{k}$ and ${\{ {\epsilon^{2}}(\vec{r})\}}_{k^
{*}}$ is the Fourier transform of $\epsilon^{2}(\vec{r})$ and its  complex conjugate
respectively, and $B(\vec{k}/k)$ is an appropriate kernel.  
Further, noting
that the favourable directions of growth of the twins are
along $[11]$ and $[1\bar1]$ directions, we find that the
 functional dependence of the kernel on $k_x$ and $k_y$ takes on a relatively
simple form. 
 We also use the fact that $[10]$ and
$[01]$ are unfavourable  directions for the growth. A simple choice of
$B(\vec k)$ having these features is
$B({\vec{k}}/k)= -{{1}\over{2}}\beta \theta(k-\Lambda){{\hat{k}}^{2}}_{x}
{{\hat{k}}^{2}}_{y}$, where ${\hat{k}}_{x}$ and ${\hat{k}}_{y}$
are the unit vectors in the  $x$ and $y$ directions, and $\beta$ is the strength
of interaction. The step function $\theta(k-\Lambda)$ has been
introduced to impose a cutoff on the range of the interaction.(See Ref. \cite{comment2}.)
Equation 2 also has the property that as the transformation proceeds,
the growth of the domains transverse to $[11]$ and $[1\bar1]$ directions are hindered.
Indeed, 
the geometrical picture of the kernel in real space is similar to
Kartha et al \cite{Kartha}.

Acoustic emission during the transformation suggests that
inertial effects are important \cite{Bales}.
This is included through the kinetic energy
of the displacement fields
\begin{equation}
T=\int d\vec{r}\rho\bigg[\bigg({{\partial u_{x}(\vec{r},t)}\over{\partial t}}
\bigg)^{2}+
\bigg({{\partial u_{y}(\vec{r},t)}\over{\partial t}}\bigg)^{2}\bigg ],
\end{equation}
where $\rho$ is the mass density. Experiments show that the AE activity depends
on the acceleration of the microplates \cite{Amen}. Since, the parent-product interface moves in the parent phase, it is
associated with dissipation which we represent by the Rayleigh dissipative functional
 \cite{Landau}.
Again, we represent
the dissipative functional entirely in terms of the deviatoric strains \cite{Bales}.
\begin{equation}
R={{1}\over{2}}\gamma\int d\vec{r}{\big({{\partial }\over{\partial t}}\epsilon
(\vec{r},t)\big)}^{2}.
\end{equation}
This is consistent with the fact that shear and bulk strains are known to equilibrate rapidly
and hence do not contribute. The equations of motion for the displacement fields are calculated using the Lagrangian $ L=T-F $
where $ F $ is the total free energy \cite{Bales} through
\begin{equation}
{{d}\over{dt}}\bigg({{\delta L}\over{\delta {\dot{u}_i}}}\bigg)-
{{\delta L}\over{\delta u_{i}}}=-{ {\delta R}\over {\delta \dot{u}_i}}, \, \, i
= x, y.
\end{equation}
\noindent
After scaling out $\rho$ and $\alpha$, the equation of motion for the deviatoric
strain $\epsilon(\vec{r},t)$ is given by
\begin{eqnarray}
\nonumber
{{{\partial}^{2} }\over{{\partial t}^{2}}}\epsilon(\vec{r},t)& = &
{\nabla}^2\bigg[{{\partial f(\vec{r},t)}\over{\partial \epsilon(\vec{r},t)}}-
\sigma(\vec{r}) - {\nabla}^2\epsilon(\vec{r},t)+
\gamma {{\partial }\over{\partial t}}\epsilon(\vec{r},t) \\
& + &2\epsilon(\vec{r},t)\int d\vec{k}B(\vec {k}/k)\{\epsilon^{2}(\vec{k},t)\}_{k}
 e^{i \vec {k} .\vec {r}} \bigg],
\end{eqnarray}

\noindent
where $\beta$ and $\gamma$ now refer to the rescaled parameters.
 There are three adjustable parameters in the model,
namely, the scaled temperature $\tau$, the strength of the long-range
interaction $\beta$ and that of dissipation $\gamma$. The effect of increasing
$\gamma$ is to reduce the twin width and that of $\beta$ is to reduce  the
 width of the
martensite domains transverse to
the habit plane directions.

Simulations are carried out by discretizing Eq.(6) on a $N \times N$ grid
 ( ie., $\Delta x = 1$) using the Euler's scheme 
with periodic boundary conditions.
The time step used  is $\Delta t=0.002$. ( Actually, we see convergence of
solutions even for $\Delta t = 0.005$.)
A pseudo-spectral technique
is  employed to compute the long-range
term \cite{Rajeev}. Computations are carried out for several values
of parameters.

Our earlier study has shown that the model exhibits hysteresis under thermal cycling
during which the energy dissipated $R(t)$, ($=- \frac{1}{2} dE/dt)$ occurs in the form of
bursts \cite{Rajeev}. 
Using Eq.(4), we compute  $R(t)$
during repeated thermal cycling in a small
temperature interval as in experiments.
We represent  the stress
field due to the random distribution of  defects by
$\sigma(\vec{r})={\sum_j^{j_{max}}}\sigma_0(\vec{r_j})exp [-|\vec{r}-\vec{r_j}|^{2}/\zeta^{2}]$,
where $\vec r_j $ refers to the coordinates of the randomly chosen defect sites and $j_{max} $
their total number. $\sigma_0(\vec{r_j}) $ is taken to be uniformly
distributed in the interval [-0.3,0.3].
\begin{figure}[!h]
\includegraphics[height=3.5cm,width=8.0cm]{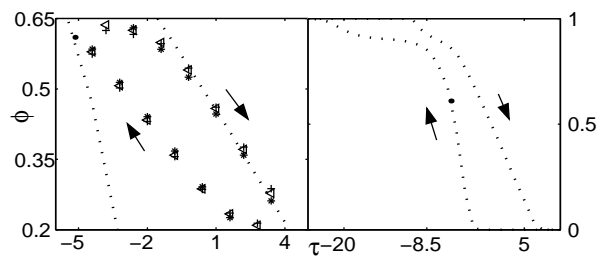}
\caption{(a) Area fraction $\phi$ during thermal cycling in the interval $\tau=$-5 to 4 for
cycles 1 ($\textasteriskcentered$) , 2 ($\triangleleft$) and 3 ($+$) .
(b) Hysteresis for the full cycle from the Austenite to Martensite phase and back.
($\bullet$) is the starting point of the cycles.}
\end{figure}
 Using the above initial
configuration, we cool the system from the austenite phase.
Then, at an appropriate point ( shown as $\bullet$ in Fig. 3b) in the full cycle,
we subject the system to
 repeated thermal cycling in a small temperature
range $\tau_{min}=-5 $ and $\tau_{max}= 4$. The starting
configuration of the martensite domains 
 for the small cycle is chosen in  a way that, it contains 
 a fair number of martensite
domains, typically, when the area fraction $\phi \approx $ 0.6.
For the cooling cycle, the final configuration  during heating cycle is taken
as the initial configuration. Calculations are  performed
for a range of parameter values of $ 50 \le \beta \le 10$ and
$5 \le \gamma \le 1$.
 As the general features remain the same,
we report here the results for
$\beta=25$ and  $\gamma=4$. Other parameter values are
$ \Lambda = 0.2,\zeta = 1$ and $N=256$.
During the first few cycles, the loops drift slightly  in the
 $\phi- \tau$ plane
which mimics the {\it training} period known in experiments,
eventually stabilizing after several cycles (Fig. 1a).
(In experiments also, the training period is typically the same
for $CuAlZn$ \cite{Lovey99}, but could be much larger in some other alloys.)
A typical plot of $\phi$ vs. $t$ for the first three cycles
is shown in Fig. 1a.

The energy dissipated $R(t)$ stabilizes after the training period. A plot of $R(t)$
for several forward and reverse cycles ( fifth to eight)  after stabilization is
shown in Fig. 2. It is clear that the energy
bursts (which mimic the AE signals), as in experiments \cite{Amen,Picor},
exhibit a near repetitive pattern in time (temperature) during successive
heating and cooling parts of the cycles.
\begin{figure}[!h]
\includegraphics[height=3.3cm,width=7.5cm]{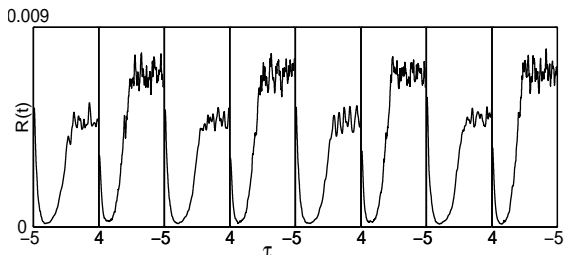}
\caption{Repetitive nature of $R(\tau)$  for cycles 5 to 8. }
\end{figure}

In order to establish a correspondence between $R(t)$ and the changes in the
spatial configuration of the martensite domains,
we have simultaneously monitored the morphology over several cycles. During
the first few cycles, the morphology changes considerably even as the
system returns to nearly the same point in the $\phi-\tau$ diagram at the
end of each cycle. Most changes occur during the first  three or four cycles as
can be seen from
Figs 3a and 3b. Figure 3a is the starting morphology for the first cycle,
 while Fig. 3b that obtained at the end of four cycles.
As can be seen, all the
curved twin interfaces have been rendered straight after going through four 
cycles.
 The snapshots of the morphology during one such stabilized
cooling and heating cycle, the fifth one, 
 at selected intervals is displayed
in Fig. 4. It is clear that during heating, the martensite plates shrink
and some even disappear. However, during cooling these domains
reappear and the eventual morphology at the end of the cycle is practically
recovered on returning to the starting point on the $(\phi,\tau)$ diagram.
As can be seen, the final configuration obtained during the fifth cycle, {\it Fig. 4d
can be seen to be practically the same as Fig. 3b}  which is the initial
configuration for the fifth cycle. Under further cycling,
the end configurations  change very little, consistent with  that observed in
experiments \cite{Lovey99}.
\begin{figure}[!t]
\mbox{
\includegraphics[height=4.0cm,width=3.7cm]{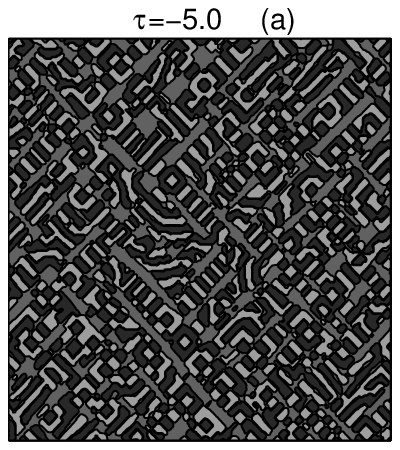}
\hspace{0.5cm}
\includegraphics[height=4.0cm,width=3.7cm]{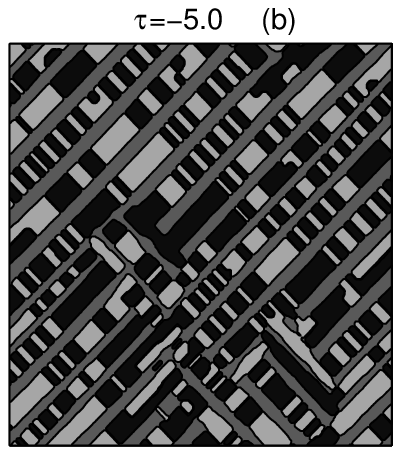}}
\caption{ Morphology of the initial configurations for (a) Cycle 1 (b) Cycle 4.
Grey cells correspond to the austenite phase, and black and white to the two martensite variants}
\end{figure}

\begin{figure}[!h]
\mbox{
\includegraphics[height=4.0cm,width=3.7cm]{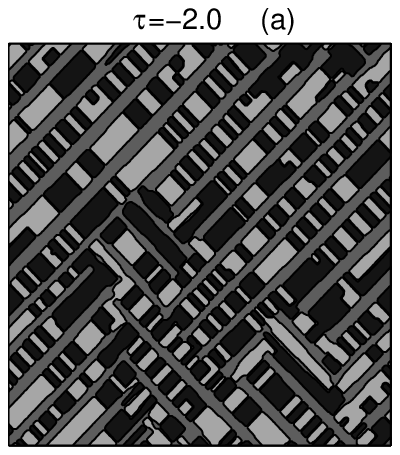}
\hspace{0.5cm}
\includegraphics[height=4.0cm,width=3.7cm]{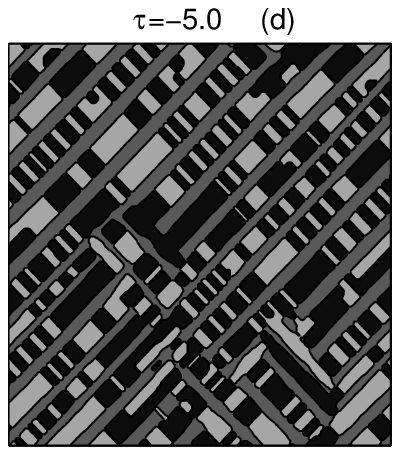}}
\mbox{
\includegraphics[height=4.0cm,width=3.7cm]{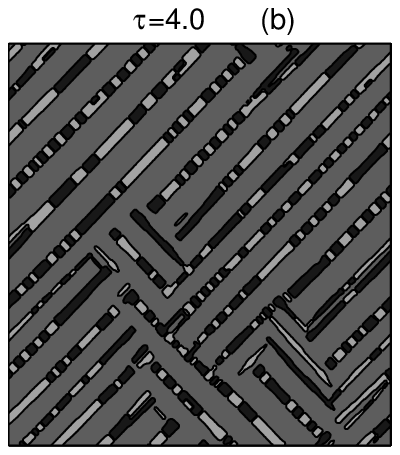}
\hspace{0.5cm}
\includegraphics[height=4.0cm,width=3.7cm]{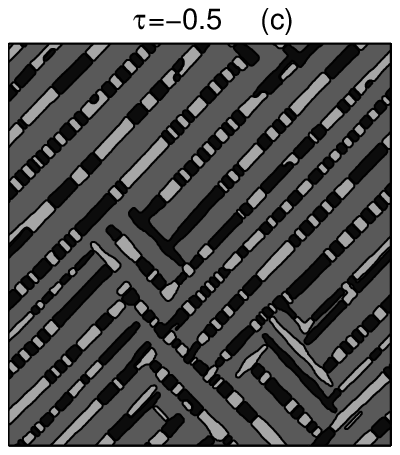}}

\caption{Sequential morphological snapshots for $\tau$ = -2.0, 4.0, -0.5 and -5.0
during the fifth cycle.
The initial configuration for the cycle is shown in Fig. 3(b).}
\end{figure}

Having established the repetitive nature of the energy bursts and the reversal of the
morphology under thermal cycling, we need to explain the role of training
cycles  in inducing this correlated behavior shown in Figs. 2 and 4. We have argued
earlier \cite{Rajeev} that the model generates a huge number of transformation induced
metastable states due to the fact that we have including threshold dynamics, slow
driving ( the rate of cooling or heating), dissipation and a fast relaxation mechanism.
The spiky nature of $R(t)$ which mimics the bursts of acoustic emission was attributed
to the following. As the temperature changes, the increase in the driving force arising
from the decrease in temperature , goes in surmounting
these local thresholds  and another part goes in to the growth  of the 
platelets, and the
rest is dissipated in the form of  bursts of energy (due to advancing  interface).
First, we note that the initial state for the repetitive small thermal cycles, Fig. 3a,
is a configuration reached by cooling from the austenite phase. Further, a
comparison of  Fig. 3a with the final morphology reached after four small thermal
cycles, Fig. 3b, reveals that the initial state Fig. 3a is a relatively shallow
metastable state. This can be seen from the fact that the sizes of the martensitic
domains of Fig. 3a,  are smaller than that in Fig. 3b (or Fig. 4d) and  the twin
interfaces of Fig. 3a are rough and considerably curved compared to Fig. 3b.
(Note  that the austenite-martensite interfaces are nearly straight). Such
configurations are generally expected to have higher energy compared to straighter ones.
The repetitive pattern of the energy bursts during successive cycles after the training
period is an indication that the system  traverses through the same set of metastable
states. During the first few cycles, the free energy landscape is so  modified that it
smoothens out the high energy barriers corresponding to the rough curved twin interfaces
in Fig. 3a with very little change in the area fraction. A crucial role in smoothening
process is actually played by the long-range interaction term, as the growth (shrinkage) of a martensite domain is influenced  by the
configuration of rest of the domains.   Computation of the free energy $F_{lr}$ arising
from the long-range interaction between the domains shows that it actually  becomes more
negative, saturating after the first few cycles. This  leads to a reduction
in the local free energy, $F_L$,  as well. The net effect is to create {\it a deeper
set of metastable states for the system to circulate for the later cycles}. Within one
such stabilized cycles, say the fifth, the starting configuration (Fig. 3b) has the
lowest free energy reaching a maximum at the end of a heating cycle
 (Fig. 4b). Thereafter it decreases during cooling. The fact that after
first few cycles, the system traverses through a deep set of free  energy  configurations
is actually reflected in Fig. 4. Note that the morphology at  $\tau = 4.0$  is similar
to that at $\tau =-0.5$ except that the widths have  increased at the expense of the
austenite regions consistent with the increase  in the well depth stated above. The
growth in the width of martensitic domains  as we decrease the temperature  is
surprisingly similar to that observed in  experiments. (See Fig. 12 of Ref.\cite{Lovey99}.)

In conclusion, our model exhibits the repetitive bursts of energy
under successive thermal cycles in a small temperature interval as
observed in experiments on acoustic emission. More importantly, these bursts
of energy are shown to be correlated to the growth and shrinkage of martensite
plates. The near full reversal of the morphology during successive cycles after
the training period can be traced to the influence of the long-range interaction term.
During the training period the long-range term  has a tendency to smoothen out
higher energy barriers in the free energy landscape. This in turn induces a
transformation pathway along a unique set {\it of low energy metastable configurations}.
As far as we know, this is the first time that the influence  of training cycles on
shape-memory effect has been elucidated.
We expect that our analysis provides a good insight into the shape memory effect
which finds immense applications in a variety of areas from mechanical actuators
to bio-medical applications \cite{Van}.
 To the  best of authors knowledge, this is the first model
which shows near full reversal of morphology under thermal cycling.


\begin{thebibliography}{19}
\bibitem{comment} Note that here we are referring to a stress free situation, but
the reversal of morphology under thermal cycling is basic to the usual shape-memory
effect induced under stress and temperature changes.
\bibitem{comment2} This choice is not unique. Further, it may be possible to have
an appropriate kernel which couples bilinear terms in $\epsilon (r)$ and $\epsilon (r^{\prime})$.
However, we expect that the functional form of $B(\vec{k}/k)$ would be complicated.
\end{thebibliography}
\end{document}